\documentclass{elsart}
\usepackage{amssymb}

\newif\ifpdf
\ifx\pdfoutput\undefined
\pdffalse 
\else
\pdfoutput=1 
\pdftrue
\fi

\ifpdf
\usepackage[pdftex]{graphicx}
\else
\usepackage{graphicx}
\fi

\ifpdf
\DeclareGraphicsExtensions{.pdf, .jpg, .tif}
\else
\DeclareGraphicsExtensions{.eps, .jpg}
\fi

\begin{document}
\begin{frontmatter}
\title{%
Coupled map gas: structure formation and
dynamics of interacting motile elements with internal dynamics}

\author{Tatsuo Shibata$^\mathrm{a,b,1}$},
\author{Kunihiko Kaneko$^\mathrm{a,2}$}
\address[kaneko]
{Department of Pure and Applied Sciences, University of Tokyo,
                Komaba, Meguro-ku, Tokyo 153-8902, Japan}
\address[shibata]
{Department of Physical Chemistry,
Fritz-Haber-Institut der Max-Planck-Gesellschaft,
Faradayweg 4-6, 14195 Berlin Germany}
\thanks[shibata-email]{present address:
 Department of Physical Chemistry,
 Fritz-Haber-Institut der Max-Planck-Gesellschaft,
 Faradayweg 4-6, 14195 Berlin Germany. e-mail:shibata@fhi-berlin.mpg.de}
\thanks[kaneko-email]{e-mail:kaneko@complex.c.u-tokyo.ac.jp}

\date{1 March 2002}

\begin{abstract}
A model of interacting motile chaotic elements is proposed.  The chaotic
elements are distributed in space and interact with each other through
interactions depending on their positions and their internal states. As
the value of a governing parameter is changed, the model exhibits
successive phase changes with novel pattern dynamics, including spatial
clustering, fusion and fission of clusters and intermittent diffusion of
elements.  We explain the manner in which the interplay between internal
dynamics and interaction leads to this behavior by employing certain
quantities characterizing diffusion, correlation, and the information
cascade of synchronization.
\end{abstract}


\end{frontmatter}

\section{Introduction}

High-dimensional dynamical systems with many interacting nonlinear
elements  have been studied extensively. Here we call such a system
``coupled dynamical system''.  Among such systems, both those with local
coupling and global coupling have been studied at length.  For instance,
in the study of pattern formation and spatiotemporal chaos, partial
differential equations and coupled map lattices~(CML)~\cite
{Kaneko1984,Kaneko1993} have been employed extensively. In a CML,
dynamical elements that are spatially distributed interact with each
other through  local interactions such as diffusive coupling. The
pattern dynamics that arise through the interplay between the internal
dynamics of individual elements and their interaction is studied by
changing the coupling strength.  Globally coupled systems with all-to-all interactions among dynamical elements also represent a typical type
of coupled dynamical system~\cite{Kaneko1990b,Shibata1998}.  They provide the
simplest type of model for complex dynamic networks.  Using globally
coupled systems, several important concepts in the study of high-dimensional dynamical systems have been developed. Such globally coupled
dynamical systems can be considered idealizations of many types of
physical, chemical and biological systems.

Coupled dynamical systems cannot always be categorized as either locally
coupled or globally coupled. In particular, living systems consist of a
huge number of dynamical elements with a variety of scales and with a
variety of types of interactions, ranging from local to global.  They
exhibit interesting collective behavior at the macroscopic level, whose
spatiotemporal scales can be much larger than that of individual
elements.  In such complex systems, even if the properties of each
element and the mechanism of its interaction with other elements are
well understood, the collective properties of the system are generally
much too complicated to be understood without extensive investigation.

In most models of coupled dynamical systems that have been employed to
this time, the interaction structure itself is fixed in the sense that
the elements with which any given element interacts as well as the
nature of the interaction are predetermined. In particular,  interacting
elements are never decoupled, and non-interacting pairs are never
coupled.

In some biological and other types of complex systems, however, the
interactions among the dynamical elements can also change in time. For
instance, consider a system of motile organisms~(e.g. bacteria).  Each
organism in this system has internal dynamics and interacts with the
others.  The interaction structure here changes in time according to the
motion of the elements. Even within a cell, interactions among molecules
change in time.  There, enzymatic reactions consist of a finite number
of molecules, each of which is a dynamical element having internal
states~\cite{Mikhailov1996}. By sharing a reactive chemical resource,
these molecules interact so as to constitute a reaction network. The
interactions among the molecules should be time dependent, due to the
motion of the molecules.  Another example is provided by neural
networks, where the coupling between elements changes in time, through
the formation and destruction of synaptic connections between neurons.

In order to describe systems of these kinds, we need to construct model
coupled dynamical systems whose couplings are dynamic, changing in a
manner that depends on the states of individual elements and the
interactions between them. In such models, we  may find a novel class of
collective behavior and self-organization.  We thus propose to study the
characteristic dynamics of systems in which the interactions between
dynamical elements change in time.

There are many possible types of models with internal
dynamics, interactions, and motile elements that exhibit interaction
structure. Instead of constructing a realistic model specific to some
phenomena, we introduce a simple abstract model possessing the above
three features.  In the present paper, we consider a coupled map whose
elements are characterized by spatial positions that change in time. By
setting a finite range for the interaction of these motile elements, the
existence of an  interaction between two given elements will change in
time. In this system, we classify several phases with distinct pattern
dynamics that depend on the parameters governing the coupling strength,
the coupling range, and the internal dynamics. The characteristic
dynamics of each phase are elucidated.

This paper is organized as follows.  In Section~2, we introduce a simple
model of interacting motile dynamical elements constituting a
combination and extension of a couple map lattice and a globally coupled
map.  The model exhibits a variety of phenomena, such as spatial
clustering, fusion and fission of the clusters, and intermittent
diffusion of the elements.  In Section~3, these phenomena are studied
for models in one- and two-dimensional space. The global phase diagram
is presented in Section~4.  Spatial clustering  is one of the
characteristic dynamical phenomena  in the present system.  In
Section~5, we study this clustering in detail for the case in which the
clusters that form are not completely isolated in space but rather
interact with each  other. Then, in Section~6, we study the case in
which each cluster is completely isolated after formation. In Section~7,
we study the case in which clusters form structures in space with a
scale larger than that of a single cluster.  These structures are
maintained by the interplay between the internal dynamics of elements
and the interactions among elements. The dynamics of the formation of
these structures are studied. The paper is concluded in Section~8 with
the summary and discussion.

\section{Couple map gas as a simple model of interacting motile
dynamical elements}

To this time, two types of coupled dynamical systems have been
investigated, one with local coupling and the other global coupling. 
Such systems consist of elements with internal dynamics and interactions
among them. In the former case, the interactions among elements are
restricted locally in space, as in the case of reaction-diffusion
equations and coupled map lattices~(CML).  In particular CMLs with local
chaotic dynamics and nearest neighbor interactions have been extensively
studied as models of spatiotemporal chaos. The simplest CML model of
this kinds is given by
\begin{equation}
 x_{n+1}(i)  = (1-\varepsilon)f(x_{n}(i)) 
 + \frac{\varepsilon}{2}\biggl(f(x_{n}(i+1))+f(x_{n}(i-1))\biggr),
\label{CML}
\end{equation}
where $n$ is a discrete time step, $i$ is the site index, and
$f(x)=1-ax^2$.
In the globally coupled case, each element interacts with all
other elements.  The simplest such model is a
globally coupled map~(GCM), given by
\begin{equation}
x_{n+1}(i) = (1-\varepsilon)f(x_{n}(i)) + 
{\varepsilon\over N}\sum_{j=1}^{N}f(x_{n}(j)),
\label{GCM}
\end{equation}
where $n$ is a discrete time step, $i$ is the element index, and $N$
is the total number of elements.  This model can be regarded as a mean-field
version of a CML, or as an infinite dimensional CML.  GCMs have also been
studied as the simplest models of complex dynamical networks.

In most studies of coupled systems carried out to this time, elements
with internal~(oscillatory) dynamics are fixed in space, and the
structure of the connections between elements is fixed in time; i.e.,
two elements that are coupled initially are always coupled.  The
coupling strengths between elements are also constant in most previously
studied models, although time dependent coupling strengths have been
introduced in some models~\cite{Kaneko1994,Ito-KK1,Ito-KK2}, with
connection topologies fixed.

In the present paper, we study a system in which the connections between
elements are time dependent in accordance with the motivation expressed
in Section~1.  For this purpose, we consider the introduction of
``motile'' elements into CML and GCM models as a natural extension of
these systems.  Two such elements interact when they are located within
a given range in space, and thus the structure of the couplings between
elements changes in time due to the motion of the elements. Because the
position of elements changes in time, two elements coupled at some time
can be decoupled at some other time.

Here, we are interested in general aspects of coupled dynamical systems
of motile elements, and for this reason we wish to consider as simple
a model as possible.  We impose the following conditions on this model:
\begin{enumerate}
\item Each element has time-dependent internal state and spatial position.

\item Each element is active in the sense that, even in isolation, its
internal state exhibits oscillatory~(chaotic) dynamics.

\item The dynamics of the internal state of a given element are affected
by local interactions between this and other elements.

\item The elements move through the influence of the force produced by
the local interaction, which depends on the internal states of the
interacting pair.
\end{enumerate}
In order to construct a model satisfying the above conditions, we
stipulate that each element has an internal state represented by a
scalar value and is characterized by a time-dependent position in real
space. For the dynamics of local internal state, a one-dimensional map
that can exhibit chaotic oscillation is employed. The dynamics of the
internal state of a given element are influenced by other elements
within a distance~${R}$ from this element.  This part of the model is
basically given in the form of coupled maps.  Furthermore, there is a
force between elements within the same range, which depends on the
internal states of the pair of interacting elements.  The motion of the
elements in space is governed by these forces. We call the class of
models defined in this manner ``coupled map gases~(CMGs)''.

A simple example of a CMG is given by
\begin{equation}
\begin{array}{lll}
x_{n+1}(i)& = &\displaystyle(1-\varepsilon)f(x_{n}(i)) 
+ \displaystyle{\varepsilon\over N_{n}(i)}\sum_{j \in {\mathcal N}(i)}f(x_{n}(j)),\\
\vec{r}_{n+1}(i)& = &
\vec{r}_{n}(i)+ {\displaystyle\sum_{j \in {\mathcal N}(i)}}
\frac{\vec{r}_{n}(j)-\vec{r}_{n}(i)}
{|\vec{r}_{n}(j)-\vec{r}_{n}(i)|}
\mathcal{F}(x_{n+1}(i), x_{n+1}(j)),
\end{array}
\label{Eq:CMG}
\end{equation}
where $x_n(i)$ is the internal state of the $i$-th element at time
step~$n$, and $\vec{r}_n(i)$ is its position in $d$-dimensional space. 
As the map governing the internal dynamics, we adopt the logistic map $f
(x)=1-ax^2$, because CMLs and GCMs employing this map have been
extensively studied as prototype models of spatiotemporal chaos and
network dynamics.  In the above equations, $|\vec{r}(j)-\vec{r}(i)|$ is
the distance between the $i$-th and $j$-th elements, the set of elements
interacting with the $i$-th element is denoted by ${\mathcal N}(i)=\{j:|
\vec{r}_{n}(j)-\vec{r}_{n}(i)|\leq{R}\}$, and the number of elements in
${\mathcal N}(i)$ is denoted by $N_{n}(i)$.  Here, 
$\mathcal{F}(x_{n+1}(i), x_{n+1}(j))$ is a ``force'', which depends on the internal states $x$ of the
$i$-th and $j$-th elements.  In the present paper, we adopt 
$\mathcal{F}(x(i), x(j))=F\cdot x(i)\cdot x(j)$ as the form of the force, using a fixed
parameter~$F$.  The elements, whose number is fixed at $N$, are located
within the fixed spatial domain $[0,L]^d$, for which we use periodic
boundary conditions.

There are four basic control parameters in our model, $a$, $\varepsilon$,
$N(\frac{R}{L})^d$ and $\frac{F}{R}$.  The role of each control
parameter is as follows: $a$~determines the nonlinearity of the local
dynamics, $\varepsilon$~is the coupling strength between elements, $N
(\frac{R}{L})^d$ is the density of elements within the interaction
range, and $\frac{F}{R}$~is the effective motility of the elements.
Throughout this paper, we fix $\frac{N{R}}{L}\sim 10$, and $\frac{F}{R}
\sim 0.01$. The effective motility is chosen to be relatively small so
that the motion of each element in space is rather smooth.

Note that the elements in the present model are ``active'' in the sense that,
even in isolation, their internal dynamics can consist of periodic or
chaotic oscillation.  This strongly contrasts with the situation in recently
studied models of the collective motion of flocks~\cite{Vicsek,Sano} and  
ants~\cite{Miramontes1993,Cole1991}, in which passive elements are employed.

In the present CMG, the motion in real space is ``overdamped'', i.e., in
the equation describing this motion, the force term is essentially
equated with the velocity, without an acceleration term included. It is
rather straightforward to introduce a momentum variable for each element
and with it include an acceleration term in the equation of motion.
Indeed we have carried out several simulations, with such equations.
However, we have found that as long as there is friction force,
the important results are unchanged. For this
reason, we consider only the overdamped case here.

The choice of the force in Eq.(\ref{Eq:CMG}) is rather arbitrary. We
have imposed the condition that the force between two elements is
attractive or repulsive, depending on the signs of the internal states
of the interacting elements. With this stipulation, the results we
obtained seem to be rather universal for any choice of $\mathcal{F}(x(i),x(j))$.
As an alternative type of force, we have also studied the case 
$\mathcal{F}(x(i),x(j))=\cos[2\pi(x(i)-x(j))]$, and we found that the qualitative
features are the same.

When the parameter~$F$ is positive, two elements interact attractively
if the signs of their internal states are the same, and repulsively if
they are opposite.  Since the variable~$x(i)$ oscillates within the
range~$[-1,1]$, the sign of the force is correlated with the degree of
synchronization of the interacting elements' oscillations.  It should be
noted, however, that interaction between two arbitrary elements is
attractive on the average if $F$ is positive, because the average of $x$
for a single logistic map (and in the present CMG) is positive for~$a<
2.0$.

As a type of coupled map system, CMG is a natural extension of a CML
in which the condition of the confinement of an element to  a single
lattice point is removed. If the elements are spontaneously located at
positions separated by a distance about~$R$, the dynamics of their
internal states are those of a CML. On the other hand, for a set of
elements located within a distance~$R$and separated from other
elements, the dynamics of their internal states are those of a GCM.
Thus the CMG spontaneously chooses the coupling structure of their
internal states ranging from those of a CML to those of a GCM.

In the case of CML or GCM, the variable $x(i)$ of each element can be
regarded as a field variable.  In the present CMG, in which  elements
move in space, internal states cannot be regarded as field variables
on fixed space.  The motion of elements in space can be regarded as
dynamics external to the dynamics of the internal state of each
element. We are concerned with the ``interplay" between such external
dynamics and the internal dynamics.

It is also interesting to consider our model from a different viewpoint.  Our
model~(in the 1-dimensional case) can be rewritten as
\begin{equation}
r_{n+1}(i) = r_{n}(i)
+ f \cdot x_{n+1}(i)\Bigl(\sum_{j \in {\mathcal N}_l(i)}x_{n+1}(j) -
\sum_{j \in {\mathcal N}_r(i)}x_{n+1}(j)\Bigr)
\end{equation}
where ${\mathcal N}_l(i)= \left\{j\,;\,j\in{\mathcal N}(i),r(j) <r
(i)\right\}$ and  ${\mathcal N}_r(i)= \left\{j\,;\,j\in{\mathcal N}(i),r(j)
>r(i)\right\}$.  In this form, the second term on the right hand side is
regarded as the gradient of the field at the position $r(i)$, which is
determined by the internal state variables~$x(i)$ of the elements
located within a distance $R$ of $r(i)$. Particles move in this chaotic
field, which they themselves form.  This interpretation of our model is
possible for any spatial dimension.

\section{Characteristic phenomena observed in a coupled map gas}\label{Sec:Pheno}

\begin{figure}
\includegraphics[width=0.32\textwidth]{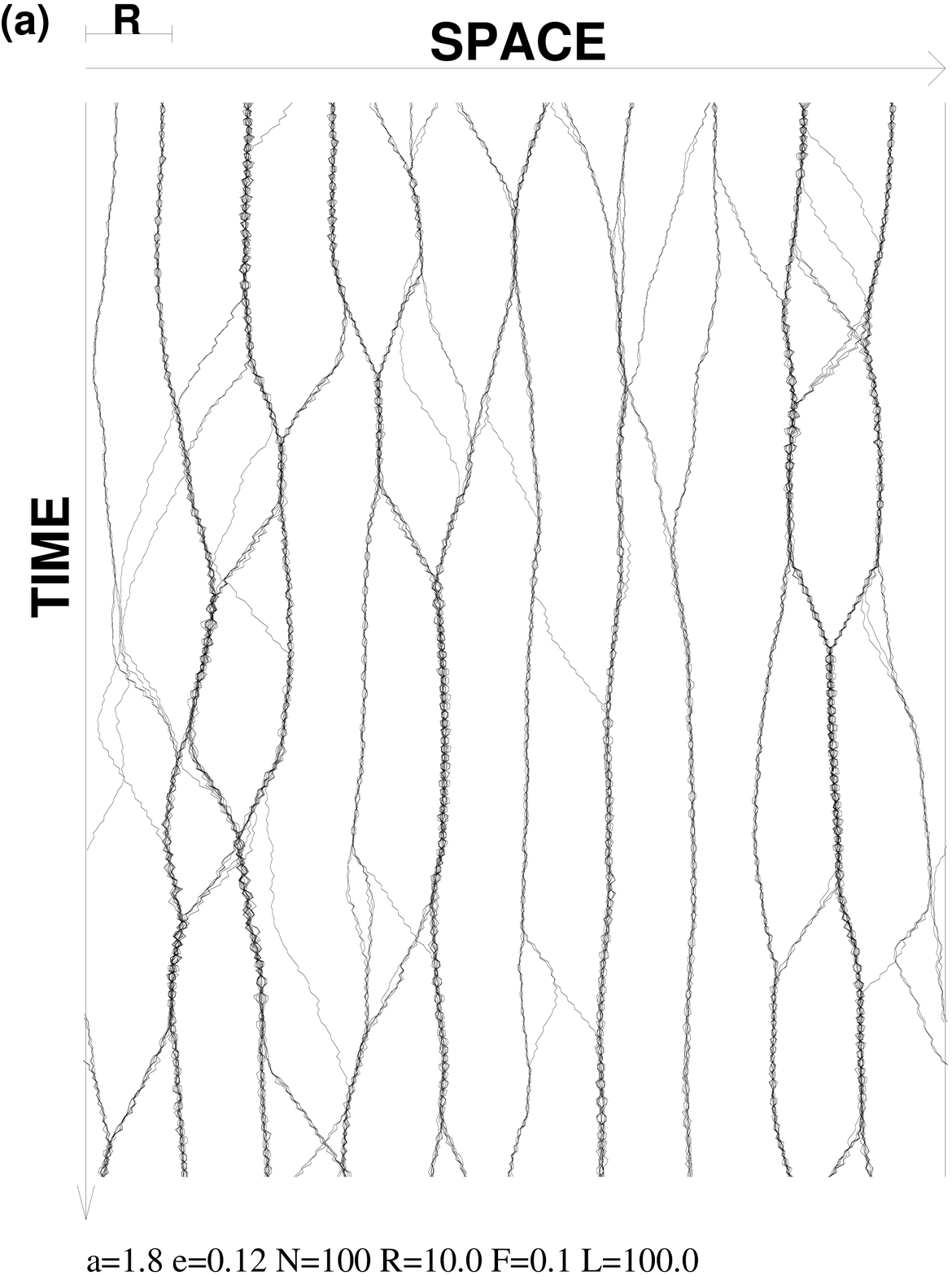}
\includegraphics[width=0.32\textwidth]{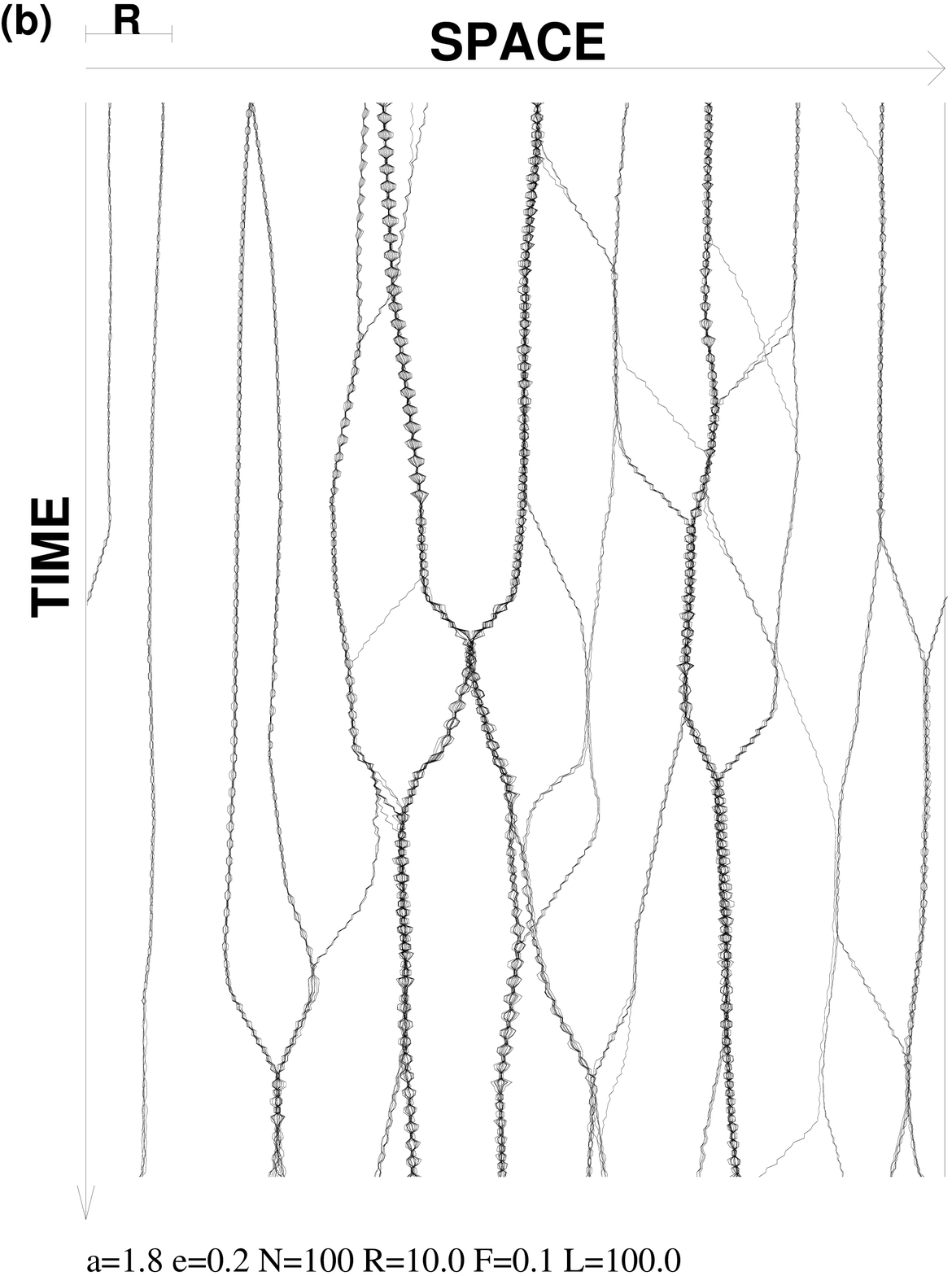}
\includegraphics[width=0.32\textwidth]{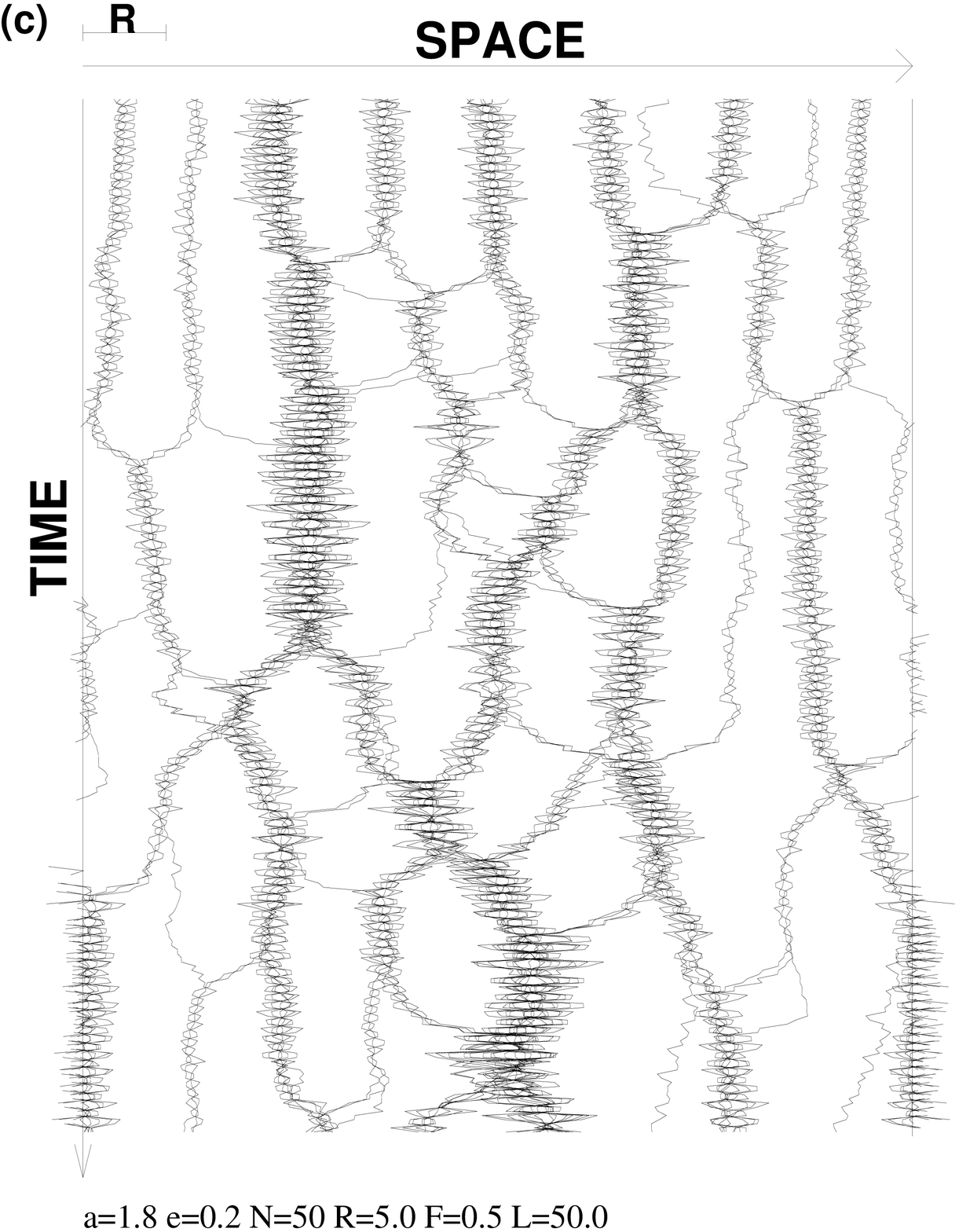}

\includegraphics[width=0.32\textwidth]{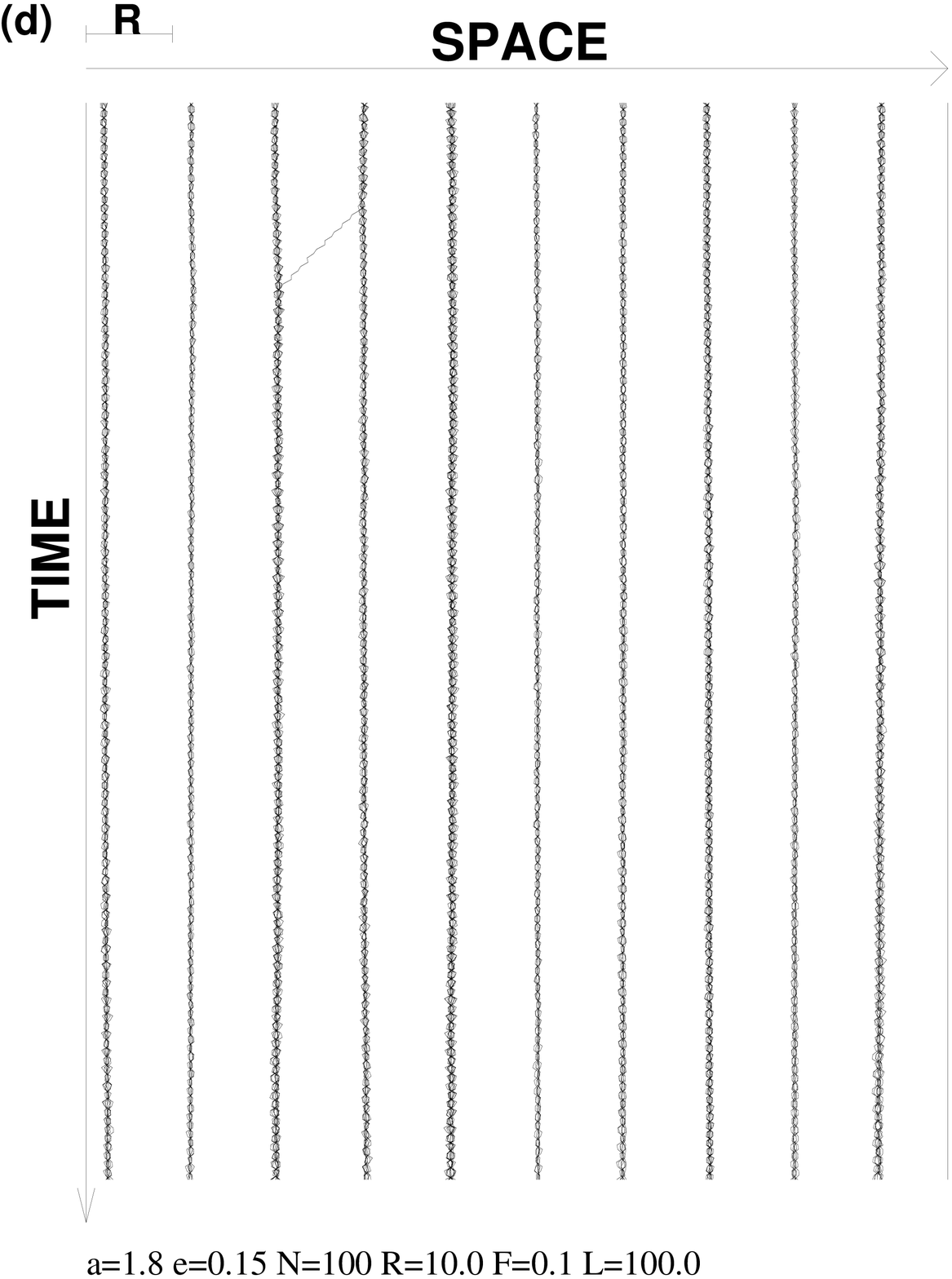}
\includegraphics[width=0.32\textwidth]{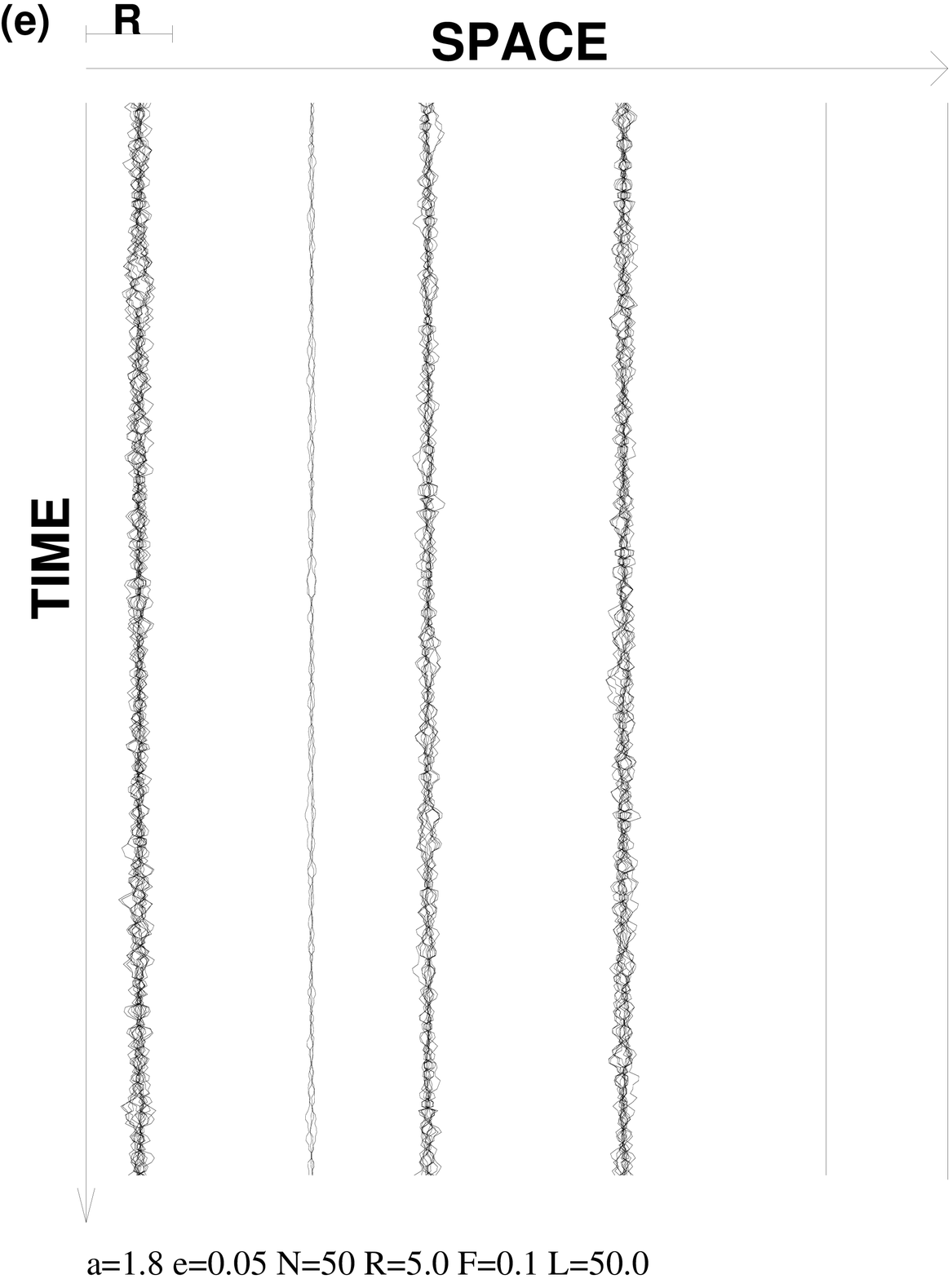}
\includegraphics[width=0.32\textwidth]{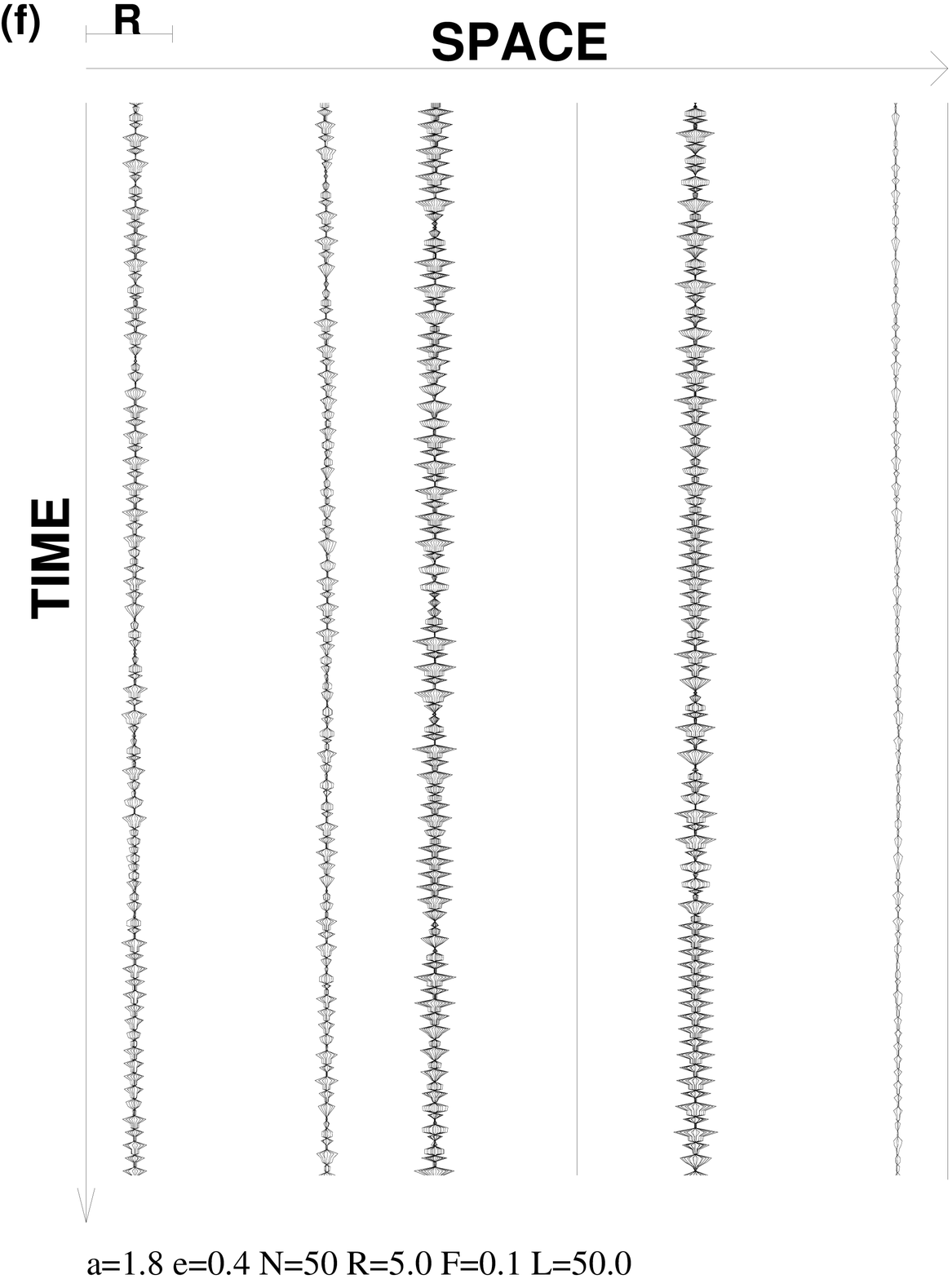}
\caption{Trajectories of elements in the presently constructed CMG in
the 1-dimensional case. The temporal evolution of the positions $r(i)$
plotted in time, given by the vertical axis. 
The trajectory of each element is plotted by
drawing  a line between successive positions at each step over
250 time steps, after transient behavior has died away.
The interval denoted by ``R'' at the top indicates the
coupling range ${R}$.\newline
(a) Fluid phase with $a=1.8$, $\varepsilon=0.12$,
$N=100$, ${R}=10.$, $F=0.1$, $L=100$,\newline
(b) Fluid phase with $a=1.8$, $\varepsilon=0.2$, $N=100$, ${R}=10.$, $F=0.1$, $L=100$,\newline
(c) Fluid phase with
$a=1.8$, $\varepsilon=0.2$, $N=50$, ${R}=5.$, $F=0.5$, $L=50$,\newline
(d) Intermittent phase with $a=1.8$, $\varepsilon=0.15$, $N=100$, ${R}=10.$,
$F=0.1$, $L=100$,\newline
(e) Desynchronized phase with $a=1.8$, $\varepsilon=0.05$,
$N=50$, ${R}=5.$, $F=0.1$, $L=50$,\newline
(f) Coherent phase with $a=1.8$,
$\varepsilon=0.4$, $N=50$, ${R}=5.$, $F=0.1$, $L=50$. 
}
\label{Fig:1D.Fluid}
\label{Fig:1D.Intermittent}
\label{Fig:1D.Coherent}
\label{Fig:1D.Desynchronized}
\end{figure}

\begin{figure}
\includegraphics[width=\textwidth]{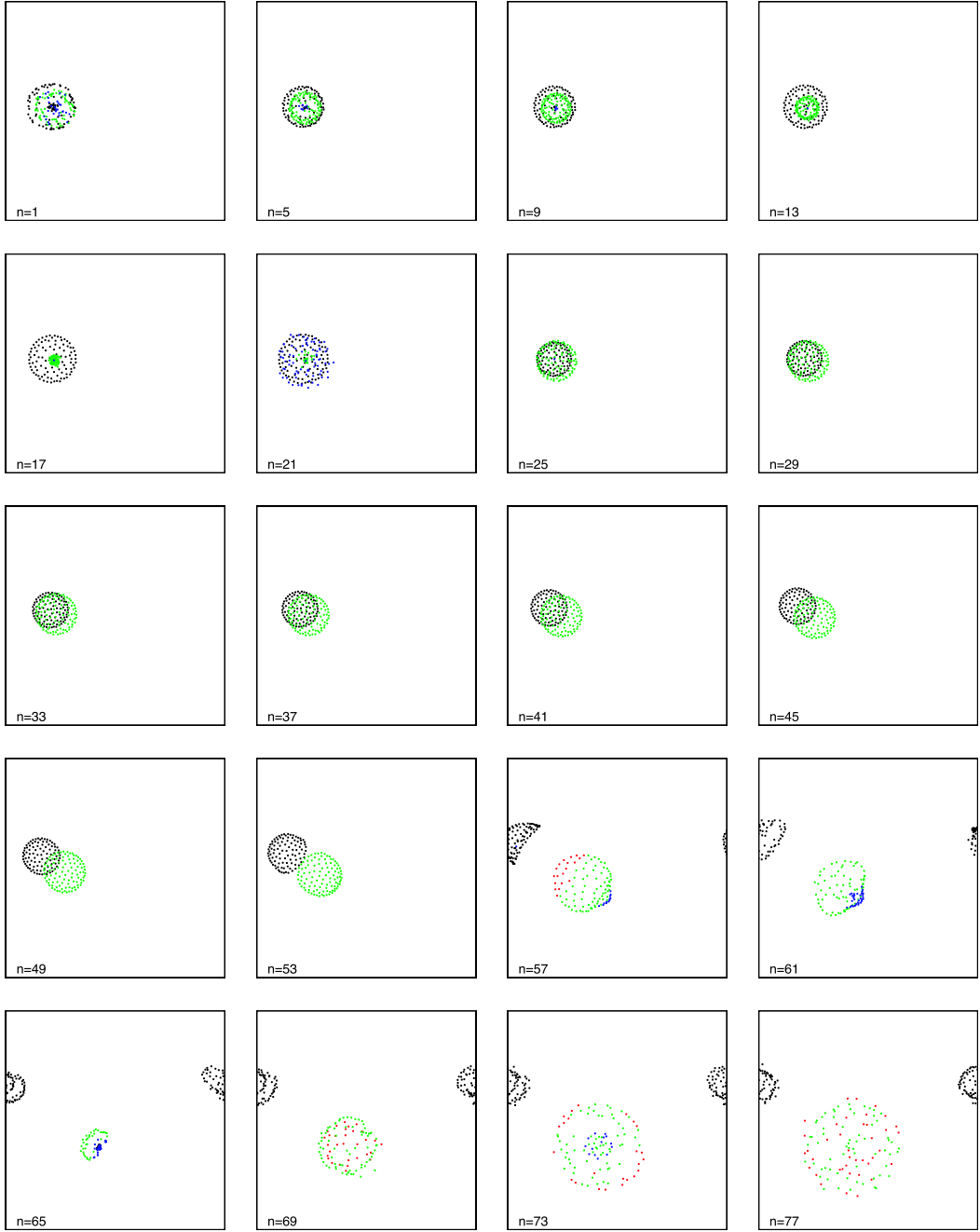}
\caption{
An example of the evolution of the present CMG in a 2-dimensional space.
Time increases from left to right and top to bottom.  Each circle
represents the position $\vec{r}(i)$ of one element.  The internal state
of each element is represented by the color of its circle.  The
snapshots displayed here are separated by four time steps where the
first snapshot~(at $n=1$) corresponds to the time at which a cluster was
formed, beginning from random initial conditions.  Periodic boundary
conditions are adopted. Here, $a = 1.8, \varepsilon = 0.3, N=200, {R} =
15.0, F = 0.2, L = 50.0.$}
\label{Fig:CMG_2D}
\end{figure}

\begin{figure}
\begin{center}
\includegraphics[width=.6\textwidth]{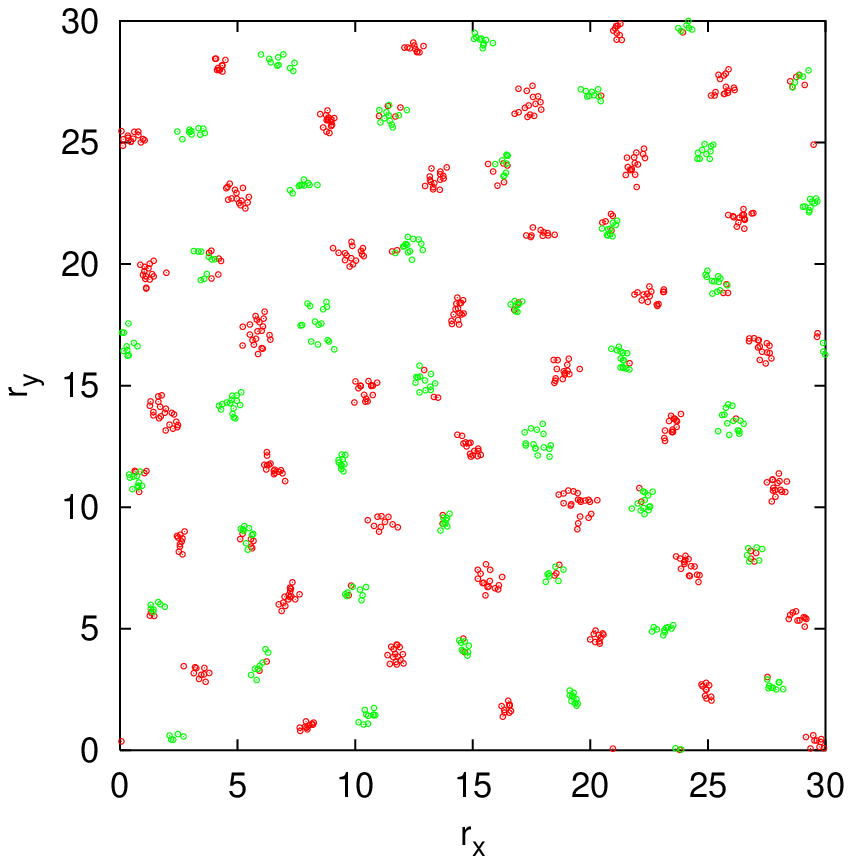}
\end{center}
\caption{A depiction of the evolution of the present CMG in a 2-
dimensional space, for the case that the value of~$F$ is negative. The
positions of elements at every two time steps are superimposed over 10
time steps; i.e., 5 snapshots are superimposed. The color indicates the
internal state of each element --red when the internal state is a
positive value and green when it is a negative value. Here, $a = 1.8,
\varepsilon = 0.2, N = 200, R = 5.0, F =-0.25, L = 30.0$}
\label{fig:repulsive}
\end{figure}

In this section, we give a rough survey of the phenomena observed in the
present model, leaving detailed analysis to later sections.
Spatiotemporal diagrams of the 1-dimensional system are given in
Figure~\ref{Fig:1D.Fluid}, where the trajectories of the elements in
real space are plotted by drawing lines between the successive positions
of the elements.

To describe the dynamics, it is useful to introduce the notion of a
``cluster'' in space.  Here, a ``cluster'' is defined as a set of
elements that are located in close proximity in such a manner that, as a
whole, they are distinguishable from the other
elements.\footnote{The term ``cluster'' is used with a different
meaning in the case of GCMs. In such models, a ``cluster'' is defined as a set of
synchronized elements}  With regard to the dynamics of clusters, we note the
following three distinct types of behavior:
\begin{enumerate}
\item
Elements gather to form a cluster whose members are not
fixed; i.e., elements enter and leave the cluster over time.
Such clusters often split into two, and two clusters often merge into one.
This formation and division of
clusters occurs repeatedly in time, as shown in
Figures~\ref{Fig:1D.Fluid}(a),~(b)~and~(c).

\item Elements form clusters that remain separated by distances
approximately equal to ${R}$ from their neighboring clusters.
Accordingly elements in each cluster can interact with those in
neighboring clusters.  In this case,  neighboring clusters exchange 
elements intermittently as shown in Figure~\ref{Fig:1D.Intermittent}(d).

\item Clusters are all separated by distances larger than~${R}$.  Hence,
there are no interactions between elements in different clusters.  In
this case, the members of each cluster are fixed.  Here, the size of
each cluster is smaller than~${R}$, so that the dynamics within each
cluster are described by a single GCM, as shown in Figures~\ref
{Fig:1D.Desynchronized}(e)~and~(f).

\end{enumerate}

This classification of the spatiotemporal dynamics displayed by our
model is also valid in the 2-dimensional case.  Examples of successive
instantaneous  patterns in the 2-dimensional case are shown in
Figure~\ref{Fig:CMG_2D}.  In a 2-dimensional system, there appears a
variety of cluster shapes, in addition to those described above.  In
this situation described by Figure~\ref{Fig:CMG_2D}, elements start to
gather to form a circular cluster.  Then, the cluster divides into two
new circular clusters. Note that the time evolution begins from random
initial conditions. The figure displays only patterns appearing after
transient behavior has died away, and after the cluster has first
formed. The formation and division of circular clusters seen here is
generally observed in the 2-dimensional system.

Another example of the 2-dimensional case is depicted in Figure~\ref
{fig:repulsive}. In this example, the value of~$F$ is negative.
Nevertheless, a lattice structure of clusters is formed. Within a
cluster, the forces between elements are repulsive, while the forces
between elements of  neighboring clusters are attractive. As far as we
have studied, the formation of clusters is generic in this class of
systems, not restricted to the present CMG.

Although the pattern formation in the 2-dimensional case is interesting,
the behavior we have found for the 2-dimensional system to this time can
conceptually be understood from the results of the 1-dimensional case.
For this reason,  we focus on the 1-dimensional case in the remainder of
the paper.

\section{Global phase diagram}
\begin{figure}[t]
\begin{center}
\includegraphics[width=\textwidth]{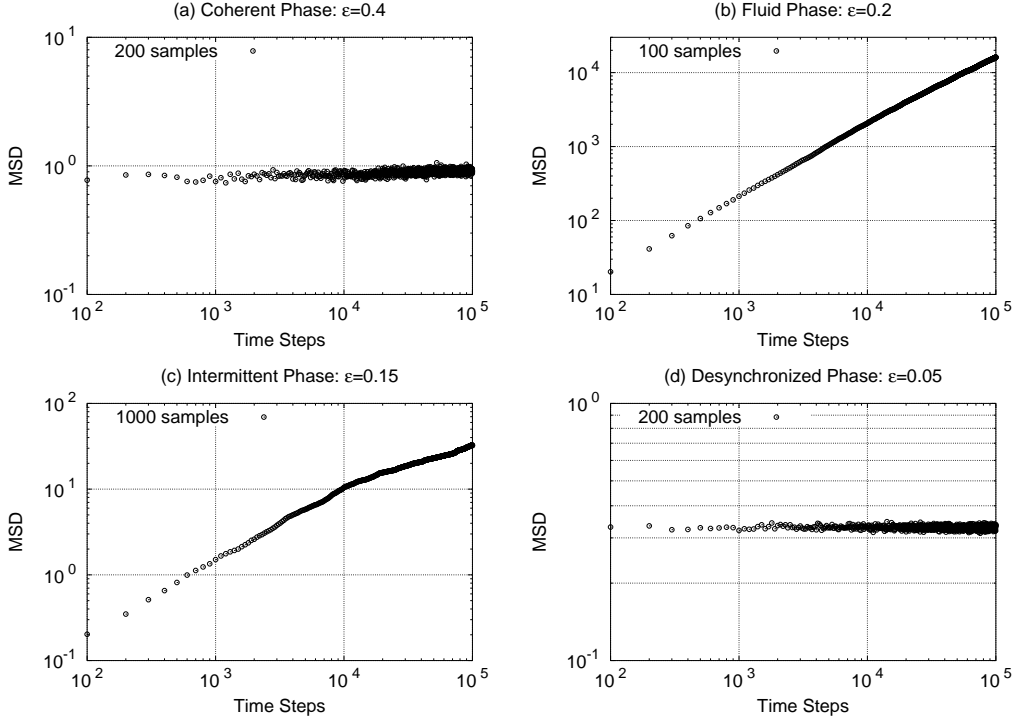}
\end{center}
\caption{Four typical evolutions of the mean square deviation~(MSD) 
$\langle\delta r^2_t\rangle = \left\langle\frac{1}{N}\sum_{i=1}^{N}
\left(r_{t+t_0}(i)-r_{t_0}(i)\right)^2\right\rangle$ of the positions of
the elements, where the average is computed over $10^5$ steps,
after discarding the initial $10^4$ steps and over the number of samples
indicated in each plot.  Here, $a=1.8$, $N=100$, ${R}=10.0$, $L= 100.0$, and
$F=0.1$.  The four types of behavior seen here correspond to four phases:
(a)~the coherent phase, (b)~the fluid phase, (c)~the intermittent
phase, and (d)~the desynchronized phase.}
\label{FIG:MSD.CMG}
\end{figure}

In this section we characterize the behavior discussed in Section~3 by
introducing some statistical measures, and also present the phase
diagram of our  model. As the measures, we introduce the following three
quantifiers, corresponding to the motion in real space, the dynamics of
the interaction structure, and the coherence of the oscillatory internal
dynamics of the elements.

\begin{enumerate}
\item  {\bf Diffusion of elements:} 
In the simulations we carried out,  elements can either move throughout
the space or be  localized in space, depending on the value of the
parameters.  In the former case, diffusive, rather than ballistic,
motion is observed.  Therefore it is useful  to introduce a quantity to
measure the diffusion of elements.  For this purpose we first define the
mean square displacement~(MSD) of the positions of elements as
\begin{equation}
\langle\delta r^2_t\rangle = \left\langle\frac{1}{N}\sum_{i=1}^{N}
\left(r_{t+t_0}(i)-r_{t_0}(i)\right)^2\right\rangle,
\end{equation}
where $\left\langle\cdot\right\rangle$ denotes the average over an
ensemble of samples. When the MSD increases linearly with $t$, it is
possible to define the diffusion coefficient of the elements,
characterizing their diffusion in real space, as
\begin{equation}
D = \lim_{t \rightarrow\infty}\frac{1}{t}\langle\delta r^2_t\rangle.
\label{EQ:diff}
\end{equation}

\item {\bf Change of coupling:}
Due to the motion of the elements, a pair of elements coupled at one time step
can be decoupled at the next time step.  The frequency that this
change in coupling takes place is a good measure for classifying 
phases with regard to the fluidity of the interaction.
As this measure, we  numerically study the conditional probability
that a pair of coupled elements remains coupled at the next
step~(denoted by $P_{CC}$) 
and the conditional probability that a pair of decoupled elements
remains decoupled at the next step~(denoted by $P_{DD}$).  In the present model,
these probabilities are given by
\begin{eqnarray}
P_{CC} = {\mathcal P}\Bigl(|r_n(i) - r_n(j)| \leq {R}\,\, \Bigl| \,\,
|r_{n-1}(i) - r_{n-1}(j)| \leq {R} \Bigr)\\
P_{DD} = {\mathcal P}\Bigl( |r_n(i) - r_n(j)| > {R}\,\, \Bigl|\,\,
|r_{n-1}(i) - r_{n-1}(j)| > {R} \Bigr)
\end{eqnarray}
for arbitrary $i$ and $j$, where ${\mathcal P}({\mathsf{A}|\mathsf{B}})$ is the
conditional probability that $\mathsf{A}$ happens given that $\mathsf{B}$ happens.
Obviously, the probabilities of the change from coupled to decoupled~
($P_{CD}$) and vice versa~($P_{DC}$) are given by$P_{CD}=1-P_{CC}$ and
$P_{DC}=1-P_{DD}$.  If there are no coupling changes,  $P_{CC}$ and $P_
{DD}$ are unity.  When there are frequent changes, $P_{CC}$ and $P_{DD}$
take small values.

\item  {\bf Coherence among the dynamics of internal states:} In
globally coupled maps, (chaotic) oscillations of all elements are
synchronized in the strong coupling regime.  Similarly, in the present
model, the internal states of elements located within the interaction
range~${R}$ can be synchronized when tightly coupled.  When the coupling
of these elements changes, such synchronization is partially or
completely destroyed.  Also, weak coherence can be sustained over all
elements, even when they are separated by distance greater~${R}$.
Therefore, to characterize the system, it is useful to introduce a
measure of the degree of coherence.
\end{enumerate}

\begin{figure}[t]
\begin{center}
\includegraphics[width=.6\textwidth]{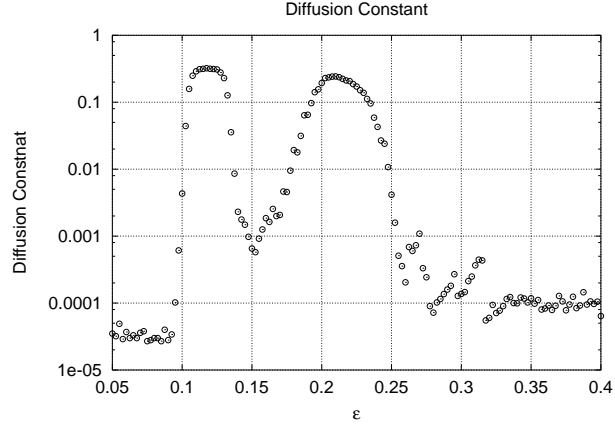}
\end{center}
\caption{Diffusion coefficient as a function of $\varepsilon$.
$\langle\delta r^2_t\rangle/t$ is plotted for $t=10000$, from 100
samples. Here, $a=1.8$, $N=100$, ${R}=10$, $L=100$, $F=0.1$.}
\label{Fig:diff_coefficient}
\end{figure}

Four typical types of evolution of the MSD with time are plotted
in~Figure~\ref{FIG:MSD.CMG}.  The diffusion coefficient of the elements
given in Eq.(\ref{EQ:diff}) calculated from the MSD is plotted as a
function of the coupling strength~$\varepsilon$ in 
Figure~\ref{Fig:diff_coefficient}, with $a=1.8$ fixed. \null From
Figure~\ref{Fig:diff_coefficient}, it can be seen that there are three
regimes in which the diffusion coefficient is very small.  These three
regimes are separated by regimes with larger values of the diffusion
coefficient. Since the diffusion coefficient plotted in 
Figure~\ref{Fig:diff_coefficient} is obtained from the MSD over a finite
time~($10^5$ steps), it is not possible to conclude that the diffusion
coefficient in these three regimes is zero, but we can reach a somewhat
certain conclusion by investigating whether the MSD increases with time
or eventually stops increasing. Examples of the time evolution of the
MSD are displayed in Figure~\ref{FIG:MSD.CMG}.

\null From the results given in Figures~\ref{FIG:MSD.CMG} 
and~\ref{Fig:diff_coefficient}, the following three types of regimes are
distinguished.

\begin{enumerate}
\renewcommand{\theenumi}{\alph{enumi}}
\item
Strong diffusion: There are two such regimes
($0.1\lesssim\varepsilon\lesssim0.13$ and
$0.2\lesssim\varepsilon\lesssim0.24$ in
Figure~\ref{Fig:diff_coefficient}), in which the time evolution of the MSD is
clearly proportional to $t$ and the diffusion constant is large.

\item
Weak diffusion: In the regime with intermediate coupling strength~
($\varepsilon\approx0.17$) among the three regimes with small diffusion
coefficient, the MSD increases almost linearly with time. Here, the
motion of the elements exhibits a diffusive behavior, whose diffusion
constant is distinctly smaller than that in the regimes with strong
diffusion.

\item
No diffusion: For the other two regimes with small values of the diffusion coefficient in
Figure~\ref{Fig:diff_coefficient}, i.e., for $\varepsilon\gtrsim0.25$
and $\varepsilon\lesssim0.1$, the MSD eventually stops increasing in
time.  In this case, the elements are localized in space.
\end{enumerate}

\begin{figure}
\begin{center}
\includegraphics[width=.6\textwidth]{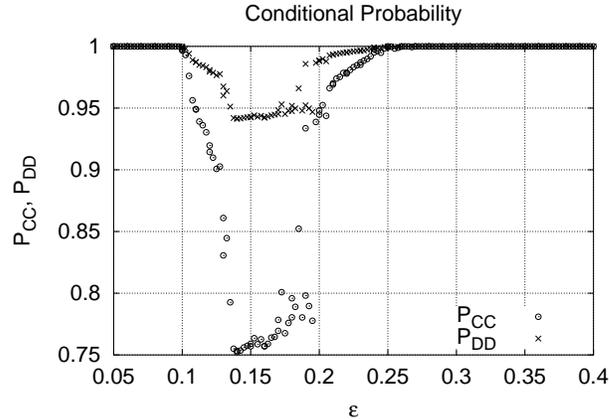}
\end{center}
\caption{The probabilities $P_{CC}$ and $P_{DD}$ defined in the text as
functions of~$\varepsilon$.  They were computed using $10^4$ time steps,
after transients.  The parameter values here  are the same as in
Figure~\ref{Fig:diff_coefficient}.}
\label{CON_PROB}
\end{figure}

Using the above classification based on the diffusion behavior, we can
classify the types of cluster dynamics discussed in the last section.
These three types of cluster dynamics are characterized as follows.
\begin{enumerate}
\item The strong diffusion regimes~(a) correspond to case~1 in the
last section (see Figures~\ref{Fig:1D.Fluid}(a), (b) and (c)), where the
formation and division of clusters continues indefinitely.

\item The weak diffusion regime~(b) corresponds to case~2 of the
last section (see Figure~\ref{Fig:1D.Intermittent}(d)), where clusters
form a lattice, and elements are exchanged intermittently between
clusters.

\item  The no diffusion regime~(c) corresponds to case~3 of
the last section (see Figures~\ref{Fig:1D.Coherent}(f)
and~(e)) where each element is localized in a
cluster, and each cluster is isolated from the other clusters.
\end{enumerate}

The probabilities $P_{CC}$ and $P_{DD}$ for the changes of couplings are
plotted in Figure~\ref{CON_PROB} as a functions of the coupling
strength.  (There, all other parameters are fixed with the same values
as in Figures~\ref{FIG:MSD.CMG}~and~\ref{Fig:diff_coefficient}.)  For
$\varepsilon\lesssim0.1$ and $\varepsilon\gtrsim0.25$, $P_{CC}=P_{DD}=1$,
since in these regimes the coupling never changes in time.  In the
intermediate coupling regime, with a small diffusion coefficient, the
probabilities are much smaller than 1.  This indicates that changes of
couplings take place frequently, even though the diffusion is much
slower.

\begin{figure}
\begin{center}
\includegraphics[height=7cm]{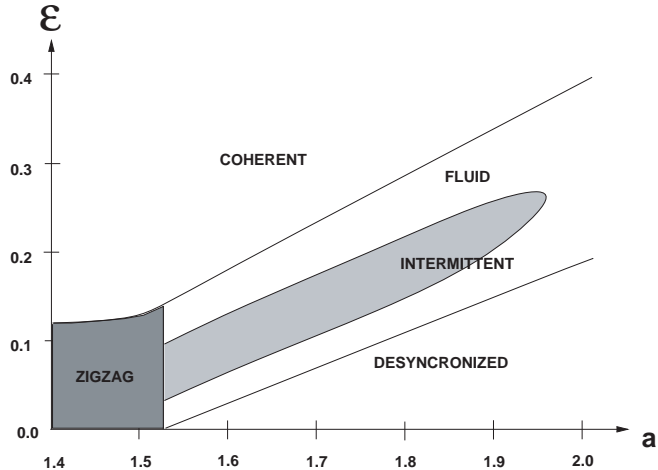}
\end{center}
\caption{The global phase diagram of CMG.}
\label{PHASE_DIAGRAM}
\end{figure}

With respect to the three types of quantities discussed above,
the MSD, the probabilities~$P_{CD}$ and $P_{DC}$ and the coherence,
we have succeeded in classifying the behavior of the present CMG
into the following five phases: 1)~coherent phase, 2)~fluid
phase, 3)~intermittent phase, 4)~desynchronized phase, 5)~checkered phase.
The phase diagram is given in Figure~\ref{PHASE_DIAGRAM} and the 
characteristic features of each phase are summarized in
Table~\ref{tabel:globalPhase}

We now describe these phases in detail.
\begin{description}
\item[Coherent phase:]
When the coupling strength is sufficiently large, elements do not
diffuse in space, as shown in Figures~\ref{FIG:MSD.CMG}(a) 
and~\ref{Fig:diff_coefficient}. In this case, the coupling among
elements does not change, as shown in Figure~\ref{CON_PROB}.  The
spatiotemporal diagram corresponding to this phase is given by
Figure~\ref{Fig:1D.Coherent}(f). Here, it is seen that elements split
into disconnected clusters separated by distances larger than the
interaction range~${R}$.  Therefore, there are no interactions between
clusters.  Within each cluster, all elements are coupled to each other.
Since all clusters are completely decoupled, the dynamics of the
elements in each cluster are reduced to those of a globally coupled map
whose total number of elements is the number of elements in the cluster.
Within each cluster, the internal oscillations of elements are
synchronized.  Thus the coherence in this phase is that of the complete
synchronization case of the GCM.

\item[Fluid phase:]
When the coupling strength is sufficiently smaller, elements diffuse
rapidly, as shown in Figures~\ref{FIG:MSD.CMG}(b)
and~\ref{Fig:diff_coefficient}.  In this case, the couplings among
elements change frequently, as shown in Figure~\ref{CON_PROB}.  In this
phase, the formation and division of clusters is repeated indefinitely,
as shown in the spatiotemporal diagram given by 
Figures~\ref{Fig:1D.Fluid}(a), (b) and (c). In this phase, the cluster
structure is dynamic, and its evolution is determined by the interplay
between the internal dynamics of interacting elements and the motion of
the elements. We discuss these dynamics in more detail in 
Section~\ref{SEC:dyn}.

\item[Intermittent phase:]
In the midst of the fluid phase, there exists a region in which 
the diffusion coefficient is distinctively small but non-zero, as shown in
Figures~\ref{FIG:MSD.CMG}(c) and~\ref{Fig:diff_coefficient}.  Even though
the diffusion here is quite weak, the coupling among elements change quite
frequently as shown in~Figure~\ref{CON_PROB}. In this case, the clusters form a lattice
structure depicted in Figure~\ref{Fig:1D.Intermittent}(d), which is
almost static with only the intermittent exchanges of elements.

\item[Desynchronized phase:]
When the coupling strength is even smaller, beyond a certain value,
the elements no longer diffuse, as shown in
Figures~\ref{FIG:MSD.CMG}(c)~and~\ref{Fig:diff_coefficient}.  In this case, the coupling
among elements also does not change, as shown in Figure~\ref{CON_PROB}.
Here, clusters are separated by distances greater than~${R}$ (as in
Figure~\ref{Fig:1D.Desynchronized}(e)) and therefore do not interact, 
while within each cluster all
elements interact with each other, and the oscillations of elements
are desynchronized.  Because all clusters are completely separated, the
internal dynamics here are those of the desynchronized state of the GCM.

\item[Checkered phase:]
There exists another phase in a region with a smaller value of the
nonlinearity parameter~$a$, in which the couplings among elements change
frequently, but elements do not diffuse in space. In this case, two
neighboring clusters are separated by a distance approximately equal
to~${R}$.  Some elements between  neighboring clusters interact with
each other at some time steps, but they can be separated by distances
greater than~${R}$ at some other times. This is the reason that here the
probabilities $P_{CC}$ and $P_{DD}$ are smaller than 1.  Within each
cluster, oscillations are strongly correlated, while oscillations of
neighboring cluster are out of phase. In this phase, elements exhibit
period-2 band oscillations. Denoting the state of~$x(i)$ only by its
sign, all elements in a given cluster oscillate as $+\rightarrow -
\rightarrow+\rightarrow-\rightarrow+\rightarrow\cdots$, while those in a
neighboring cluster oscillate perfectly out of phase as $-\rightarrow +
\rightarrow-\rightarrow +\rightarrow \cdots$.  Thus the states $x$, when
represented in this way, form a checkered spatial pattern among cluster.
\end{description}

\begin{table}
\begin{tabular}{|l||c|c|c|}
\hline
&Diffusion&
\parbox{50pt}{Change of Coupling}
&
\parbox{70pt}{Coherence among internal dynamics}
\\
\hline\hline
Coherent Phase&$\times$&$\times$&$\bigcirc$\\
\hline
Fluid Phase&$\bigcirc$&$\bigcirc$&$\times$\\
\hline
Intermittent Phase&$\bigcirc$(weak)&$\bigcirc$(frequent)&$\bigcirc$(Zigzag)\\
\hline
Desynchronized Phase&$\times$&$\times$&$\times$(Desynchronized)\\
\hline
Checkered Phase&$\times$&$\bigcirc$&$\bigcirc$(Checkered)\\
\hline
\end{tabular}
\caption{The characteristics of each phase. (See text for details.)}
\label{tabel:globalPhase}
\end{table}

\section{Formation and collapse of clusters in the fluid phase}
\label{SEC:dyn}

In a coupled dynamical system, the dynamics of elements often become
synchronized so that they come to form groups in phase space. Such
groups are often also referred to as 
``clusters''~\cite{Kaneko1990b,oscillator}.  In the present case, such
synchronization of oscillations influences the formation and collapse
of clusters in real space.  Conversely, the motion of elements in real
space influences the internal dynamics of elements.  
As a result, 
formation, fusion and fission of the
clusters occur repeatedly in the fluid phase.
Studying the interrelation between the internal dynamics and
the motion in real space, we have found that the process of cluster
formation, fission and fusion in space can be described as follows.


%

\begin{description}
\item[step 1.]
The interactions among elements tend to synchronize the internal states
of elements.  However,  chaotic nature of the internal dynamics tends
to destroy this synchronization.  When synchronization dominates, forces
between elements on the average become attractive, and clusters are
formed.

\item[step 2.]
Within each cluster, the dynamics are of a GCM type (to the extent
that the influence of neighboring clusters can be ignored). A number
of synchronized groups are formed among the elements in a cluster. 

\item[step 3.]
In some of such groups
the sign of the internal states evolve as 
$+\rightarrow-\rightarrow+\rightarrow-\rightarrow\cdots$,
while that of in other groups evolve as
$-\rightarrow+\rightarrow-\rightarrow+\rightarrow\cdots$.
Then, the force between two elements from such two types of groups is
repulsive, while the force between the elements within the same type
of group is attractive.


\item[step 4.]
Due to the repulsive forces between groups, the cluster divides into
two new clusters, if the cumulative force is sufficiently strong. This
repulsion continues until the distance between the two new clusters is
larger than ${R}$.  Consequently, the two new clusters come to be
located at a distance of about ${R}$.

\item[step 5.]
When the distance between the two new clusters becomes larger than~${R}$,
they no longer interact.  Initially, almost all elements within
one of these clusters (i.e., elements within the range ${R}$) are
nearly synchronized, but this state is unstable, because synchronized
chaos in the GCM is unstable in this parameter region.  Thus the
process returns to step 2.
Thus each cluster has the potentiality of repeated division.

\item[step 6.]
If a cluster interacts with other clusters, depending on the internal
dynamics of these clusters, these clusters can fuse together.
Then, the evolution of this newly merged cluster
returns to step~2.
\end{description}
In this way, the fusion and fission processes of clusters
continues indefinitely.

\begin{figure}
\begin{center}
\includegraphics[width=\textwidth]{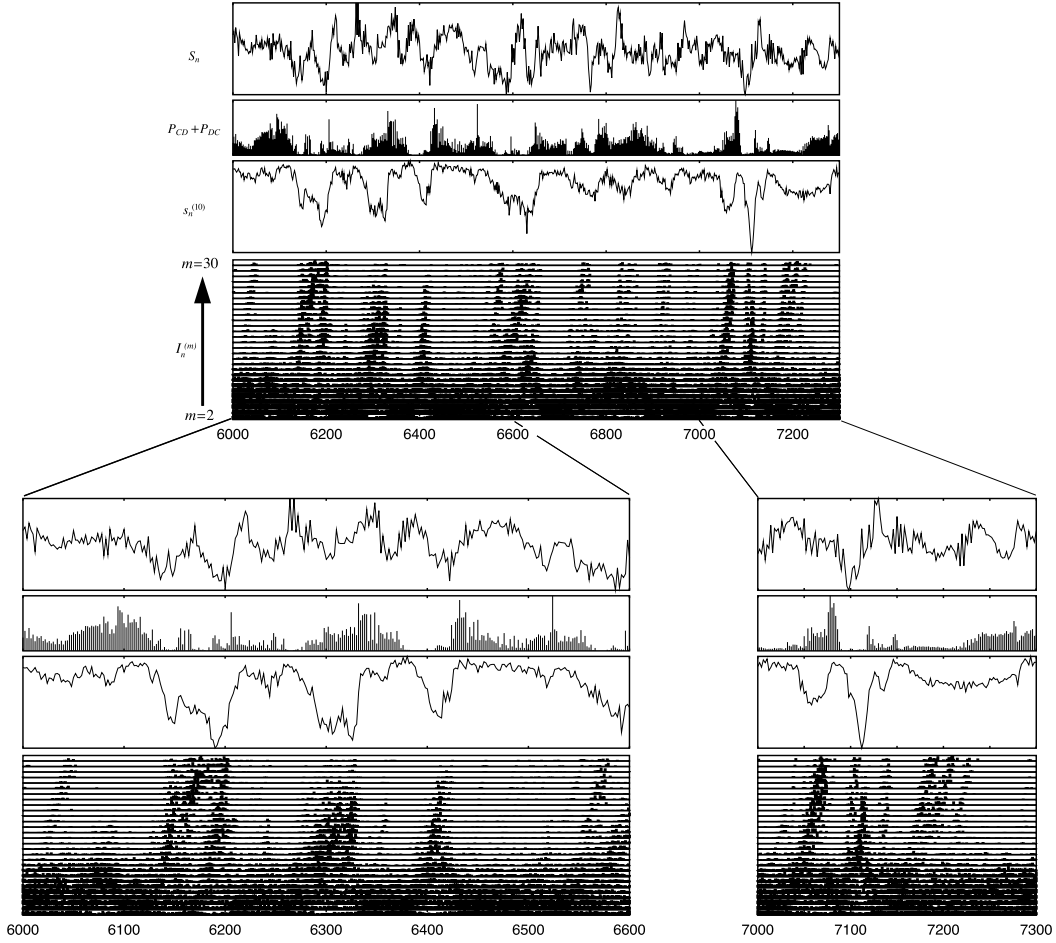}
\end{center}
\caption{The information creation per bit~$I^{(m)}_n$,
the $m$-bit entropy~$s^{(m)}_n$ for $m=10$,
the conditional probability~$P_{CD}+P_{DC}$
and the spatial entropy~${\mathcal S}_n$ plotted as functions of
time for the fluid phase.
Here, $a=1.8, \varepsilon=0.2, N=100, {R}=10., F=0.1, L=100.$
}
\label{Fig:picc}
\end{figure}

In order to understand the above described process in a more
quantifiable manner, we now characterize the behavior of the spatial
motion and the internal degrees of freedom quantitatively.

In the fluid phase, one of the most important processes in the
internal dynamics is the synchronization.  In the synchronization
process, the difference between the internal variables of two elements 
decreases with time. This process can
be effectively seen by measuring the information creation per
bit~$I^{(m)}_n$, introduced in Ref.\cite{Kaneko-Info-Cascade}, where
$n$ is a discreate time step and $m$ is a particular integer value.
For this, we partiton the space of the internal state into intervals
of a size~$2^{-m}$.
Then, we define the distribution function
$P^{(m)}_n(j),\quad(j=0,\cdots,2^{m+1}-1)$ as the fraction of elements
whose internal states take values in the interval
$[j\times 2^{-m}-1,(j+1)\times 2^{-m}-1]\quad(j=0,\cdots,2^{m+1}-1)$
at time step~$n$.  Here, we disregard the position of each element.
Then, the information creation per bit~$I^{(m)}_n$ at
time step~$n$ is defined by
\begin{equation}
I^{(m)}_n=s^{(m)}_n-s^{(m-1)}_n,
\end{equation}
where $s^{(m)}_n$ is the $m$-bit entropy at time step~$n$ given by
\begin{equation}
s^{(m)}_n=-\sum_{j=0}^{2\times2^m-1}P^{(m)}_n(j)\log{P^{(m)}_n(j)}.
\end{equation}
Since the $m$-bit entropy is a nondecreasing function of $m$,
$I^{(m)}$ cannot take a negative value.
If two elements are synchronized to a precision of $2^{-\ell+1}$,
the value of $s^{(m)}$ increases at $m=\ell$.
Therefore $I^{(m)}$ takes a positive value when $m=\ell$,
while $I^{(m)}$ equals zero for values of $m$ less than $\ell$.
If the two elements are becoming synchronized,
the largest value $m=\ell$ at which $s^{(m)}$ increases 
and $I^{(m)}$ is non-zero increases with time.
Thus, the synchronization process can be viewed as the increase of the largest value of $m$ for which $I^{(m)}$ is greater than zero.

At the bottom of Figure~\ref{Fig:picc}, $I^{(m)}_n$ for $\quad m=2,\cdots,30$
is plotted as a function of time.  Several synchronization processes
can be seen in this figure, as the increase of the largest value of
$m$ for which $I^{(m)}_n>0$.
In Figure~\ref{Fig:picc}, the time evolution of the $m$-bit entropy
$s^{(m)}_n$ is also plotted for $m=10$.
This entropy measures the degree to which elements 
are synchronized.  The synchronization processes can also be 
identified by the decrease of~$s^{(10)}_n$,
while once synchronization has been established, its breakdown can be
identified by the increase of~$s^{(10)}_n$.

Change in the spatial configuration of elements is
characterized by the spatial entropy.  First, we partition
the entire space into intervals of a given size.  Then we define the
the fraction of elements in each interval~$j$
as~${Q_n(j)}$. Here, the width of an interval is chosen to be approximately the same as the effective width of the cluster~(see the caption of
Figure~\ref{FIG:potential}). In our simulations, we chose the value $L/64$.
The spatial entropy is computed as
\begin{equation}
{\mathcal S}_n=-\sum_{j=1}^{}Q_n(j)\log{Q_n(j)}.
\end{equation}
\null~From the time evolution of ${\mathcal S}_n$ displaying in
Figure~\ref{Fig:picc}, one can see that ${\mathcal S}_n$
decreases in the process of cluster formation and increases in the
process of cluster collapse.

In Figure~\ref{Fig:picc}, ${\mathcal S}_n$ is plotted as a function of
time. As the synchronization proceeds, as is shown by the increase of
the largest $m$ for which~$I^{(m)}>0$, the decrease of the spatial
entropy~${\mathcal S}_n$ is observed. This indicates cluster formation
processes. After the synchronization process becomes completed, the
spatial entropy~${\mathcal S}_n$ stops decreasing.  Then, ${\mathcal S}_n$
begins to increase, indicating that the process of cluster
collapse has begun.  The collapse of cluster accompanies the breakdown
of the synchronization among the internal dynamics of its element. As
a result, the $m$-bit entropy~$s^{(10)}_n$ also begins to increase. In
Figure~\ref{Fig:picc}, the conditional probability~$P_{CD}+P_{DC}$ is
plotted as a function of time, indicating the frequency of coupling
change~(see Section~4).\null~From this, it is seen that the collapse
of a cluster leads to an increase in the frequency of coupling change.

In this way, the formation and collapse of clusters take place
repeatedly due to the interplay between the internal dynamics and the
spatial motion of the elements.  Such processes are expected to take
place at any point in  space.  Although $I^{(m)}_n$, $s^{(m)}_n$,
$P_{CD}+P_{DC}$ and ${\mathcal S}_n$ are computed by treating all
elements identically without taking account of spatial structures, we
can clearly see the processes of cluster formation and collapse
reflected in the behavior of  these averaged quantities.

\section{Absence of fusion and fission of clusters in the coherent and
desynchronized phases}
\label{SEC:dyn2}

In this section, we study cases in which  neither fusion nor fission of
clusters takes place once clusters are formed. One such case is in the
coherent phase with the coupling strength sufficiently large, and the
other is in the desynchronized phase with the coupling strength
sufficiently small.

In the case of the coherent phase, elements in a cluster interact with
each other strongly, so that the internal dynamics of the elements in
the same cluster are highly synchronized.  It is the attraction among
elements resulting from synchronization that causes a cluster to form.
For a system in the synchronized state, beginning from random initial
conditions, after transient behavior dies away, clusters are formed and
exist isolated in space, with  no interaction among them.  Hence, the
internal dynamics of the elements in any given cluster are those of a
GCM in its coherent state.  If coherent oscillation of these isolated
GCM systems is stable, this stable cluster formation is guaranteed, and
the resulting clusters will be separated by a distance greater than~${R}$.

In the case of the desynchronized phase, elements are coupled weakly,
so that there is no synchronization, and the coherence among the
internal dynamics of elements is weak.  Even in such a case, the
elements form clusters that are also separated by distances greater
than ${R}$, and again there exists no interaction among clusters. In
this case, within a cluster,  the internal dynamics of the elements
are those of the desynchronized state of a GCM. Then, the forces
between elements change sign frequently, and the interactions between
elements can often be repulsive. Thus the reason that no elements
escape from a cluster is not immediately evident.

A possible reason for this fact may be that the sum of forces among
the elements gives an attraction towards the center of the cluster,
even if each two-body force is repulsive.  In order to check if this
could be the case, the direction and strength of the force acting on
each element was calculated as a function of the distance from the
center of mass of the cluster. The strength of the force was
calculated as follows. The strength of the force acting on an element
is given by the second term on the right hand side of Eq.(\ref{Eq:CMG}).
Then, the position of an element in the cluster is defined
by its distance from the center of the cluster.  Here, the center of a
cluster is defined as the center of mass  of the elements that are
located within a distance~$d$ of a given element at a given time step.
Since even elements that directly interact with each other can be
members of different clusters, $d$ should be smaller than~${R}$.  The
strength of the force on an element and its fluctuation as functions
of the position can depend on the distance~$d$.  However, if a 
well-defined cluster exists, there should be a range of values of $d$ for
which the value of the force calculated in this manner does not
strongly depend on~$d$. We take the value of~$d$ within such an
interval. The position of the center of a cluster computed in this
way, by choosing an arbitrary element, is approximately the same for
any chosen element within the cluster.

\begin{figure}
\begin{center}
\includegraphics[width=\textwidth]{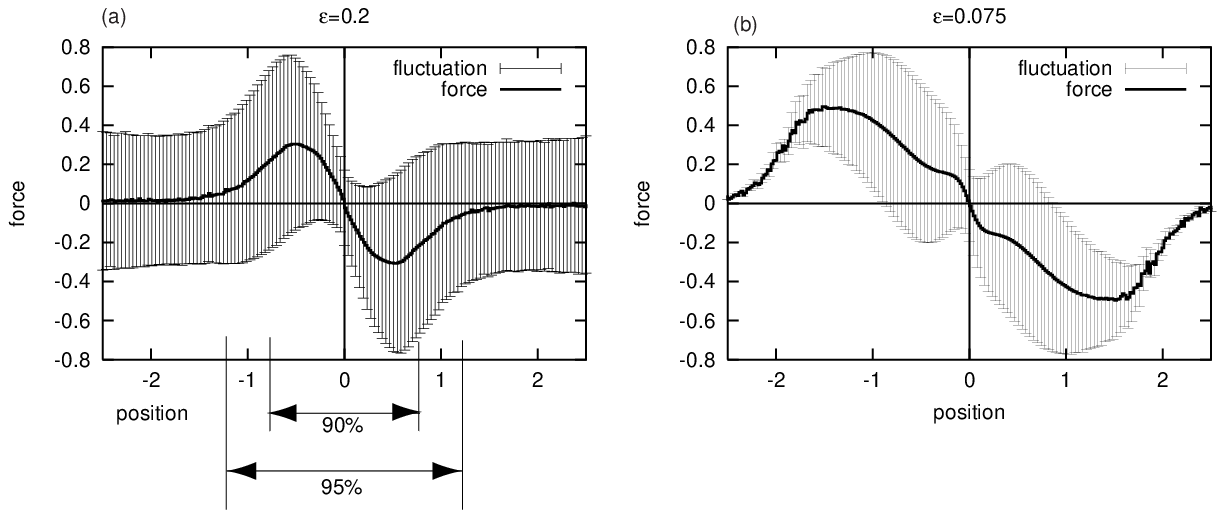}
\includegraphics[width=\textwidth]{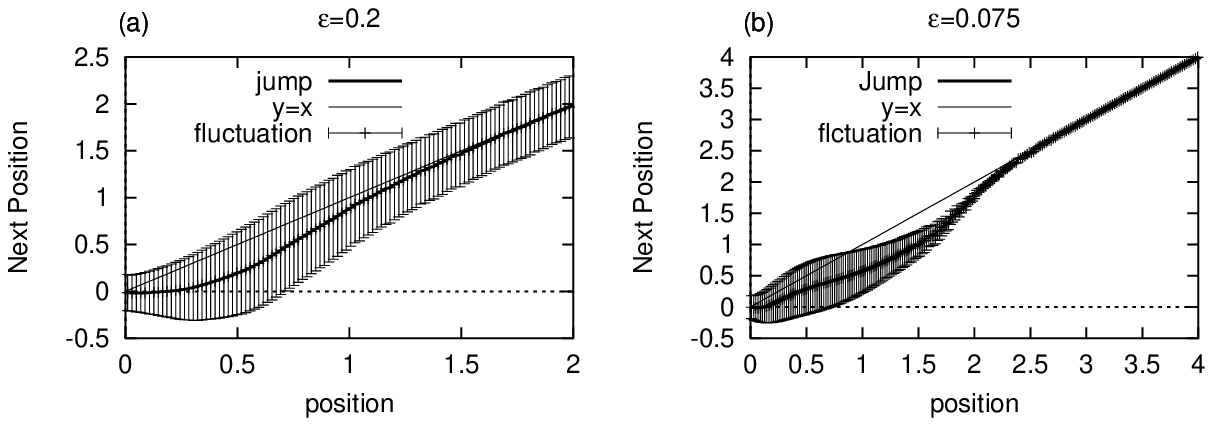}
\end{center}
\caption{(a)~and~(b):  The strength of the force acting on an element and its
fluctuation as functions of the position of the element
in the cluster. In the case of (a), elements stay within the range
$[-0.77,0.77]$ with probability~$90\%$, and within the range 
$[-1.22,1.22]$ with probability~$95\%$. We can consider the lengths of these intervals are roughly the effective width of
the clusters. (c)~and~(d): The position of an element at one time
step as a result of the force experienced at the previous time step 
as a function of the
position. (For the method of calculation, see the text.)  The average
strength of the force and its fluctuation were computed over $10^4$
time steps, after discarding the initial $5000$ steps, and averaged over
$100$~samples, starting from random initial conditions. Here, $a=1.8,
N=100, {R}=10.0, F=0.1, L=100.0$, (a)~$\varepsilon=0.2$,
$d=4.0$, (b)~$\varepsilon=0.075$, $d=5.0$, (c)~$\varepsilon=0.2$,
$d=4.0$, (d)~$\varepsilon=0.075$, $d=5.0$.}
\label{FIG:potential}
\end{figure}

In Figures~\ref{FIG:potential}(a)~and~(b), the strength of the force experienced by an element and
its fluctuation are plotted as functions of the position, 
and in Figure~\ref{FIG:potential}(c)~and~(d), the
position of an element at one time step as a result of the force
experienced at the previous time step is plotted as a function of the
positions.  In both the cases of the
fluid phase, in~(a)~and~(c), and the desynchronized phase in~(b)~and~(d),
elements around the center of the cluster are attracted to the center, on
average. However, by taking the fluctuations into account, these two
phases are clearly distinguished.

For the fluid phase,  the fluctuation~(indicated by the error bars in
the figure) always crosses over the~line~$y=x$ in 
Figure~\ref{FIG:potential}(c). It is thus seen that elements can
occasionally escape from the cluster. For the desynchronized phase, by
contrast, the fluctuation does not cross over the~line~$y=x$ for
positions within a particular range near when attractive force
disappears. This indicates that in the desynchronized phase, the
elements around the edge of the cluster are with certainly attracted
to the center, and therefore elements cannot escape from the cluster.
We thus understand that the collective attraction in the
desynchronized phase is the source of cluster formation.

\section{Structure formation and weak diffusion in the intermittent
phase}\label{SEC:STR}

\begin{figure}
\begin{center}
\includegraphics[width=\textwidth]{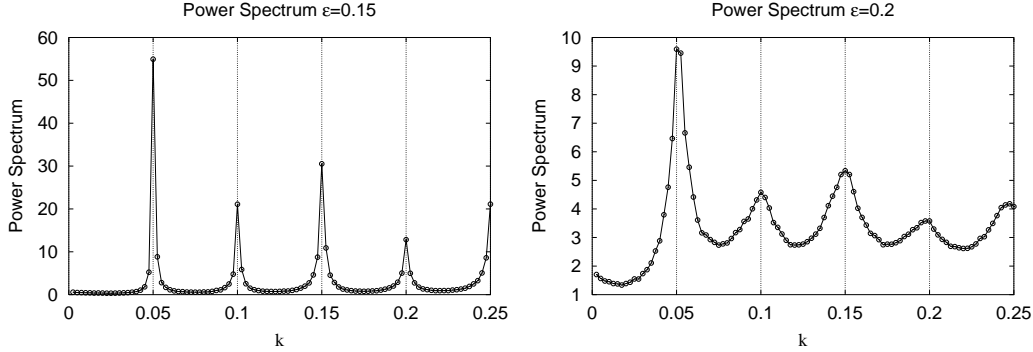}
\end{center}
\caption{Power spectrum, 
$P(k)=\left\langle\frac{1}{N}
\sum_{j=1}^{N}x(j)e^{2\pi\mathrm{i}kr(j)}\right\rangle$.  
The vertical lines in the figure are at intervals of $\frac{1}{2{R}}$.
The average is computed over $10^4$ time steps, after discarding the
initial $10^4$ steps and averaged over $10$~samples  starting from
random initial conditions. Here, $a=1.8, N=400,{R}=10.0, F=0.1, L=
400.0$, 
(a)~intermittent phase $\varepsilon=0.15$
(b)~fluid phase $\varepsilon=0.2$.}
\label{FIG:power.cmg}
\end{figure}

In this section we study the intermittent phase by considering
structure formation and intermittent behavior.

In the intermittent phase, after transient behavior has died out,
through the repeated fusion and fission of clusters, clusters
eventually form a lattice structure with a nearly fixed
spacing~${R}$~(see Figure~\ref{Fig:1D.Intermittent}(d)).
This lattice structure can be clearly characterized by the power
spectrum defined as
\begin{equation}
P(k)=\left\langle\frac{1}{N}\sum_{j=1}^{N}x(j)e^{2\pi\mathrm{i}k r(j)}\right\rangle,
\end{equation}
in which the internal degrees of freedom are also taken into account.
Figure~\ref{FIG:power.cmg} displays typical power spectra for the
intermittent phase and the fluid phase.  The sharp peaks in
Figure~\ref{FIG:power.cmg}(a) indicate the formation of lattice structure
over a long spatial distance.  The first peak is located at the wave
number~$\frac{1}{2{R}}$ not~$\frac{1}{R}$, implying some structure
with wavelength $2{R}$.  This is because the internal degrees of
freedom of the elements in a cluster are approximately in phase,
whereas those of the elements of neighboring clusters are  of nearly
opposite phase.
Hence the clusters form a checkered pattern, and the wavelength 
is~${2R}$.  Even in the case of the fluid phase, depicted in
Figure~\ref{FIG:power.cmg}(b), there is a peak at around the wave
number $\frac{1}{2{R}}$. This peak is not as sharp here as it is in
the case of the intermittent phase, due to the successive formation
and collapse of clusters.

\begin{figure}
\begin{center}
\includegraphics[width=0.6\textwidth]{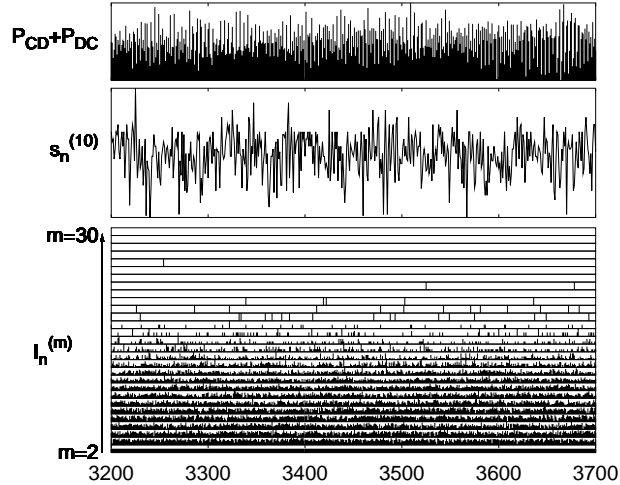}
\end{center}
\caption{The information creation per bit~$I^{(m)}_n$,
the $m$-bit entropy~$s^{(m)}_n$ for $m=10$,
the conditional probability~$P_{CD}+P_{DC}$,
and the spatial entropy~${\mathcal S}_n$ as functions of
time for the intermittent phase.
Here, $a=1.8, \varepsilon=0.15, N=100, {R}=10., F=0.1, L=100.$}
\label{FIG:piccIntermittent}
\end{figure}

In the fluid phase, synchronization among elements leads to the
formation of clusters, as shown in Section~5, where the
synchronization process was characterized by the information creation
per bit,~$I^{(m)}_n$. In the intermittent phase, the internal dynamics
of elements within a cluster are roughly in phase.  In contrast to the
fluid phase,  these dynamics do not exhibit any symptoms of
synchronization among elements. In Figure~\ref{FIG:piccIntermittent},
$I^{(m)}_n$ is plotted as a function of time. No increase of the
largest value of $m$ for which~$I^{(m)}_n>0$ is observed. This
indicates that no synchronization process takes place.

If one cluster is isolated from the lattice structure, in phase motion
within the cluster cannot be maintained  and as a result, the cluster
divides into two. Thus, in the intermittent phase, the interactions
among clusters are essential for the formation of clusters.  As shown
in Figure~\ref{CON_PROB}, the high activity of interaction between
clusters also suggests the importance of interactions.

In spite of the absence of synchronization among elements and the
frequent change of the interactions among clusters, the diffusion constant
in the intermittent phase is extremely small but not zero,
as shown in Figure~\ref{Fig:diff_coefficient}. This
indicates that the structure is stable most of the time.  Then, we are faced with the question of why this structure is so stable.

\begin{figure}
\begin{center}
\includegraphics[width=\textwidth]{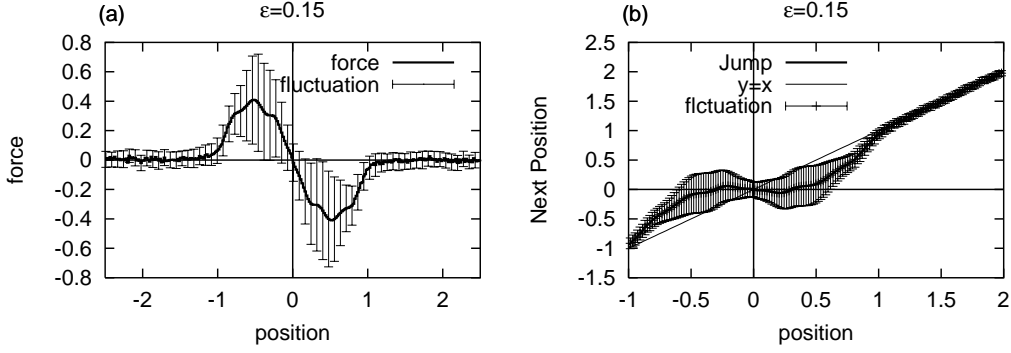}
\end{center}
\caption{(a) The strength of the force acting on an element and its
fluctuations as functions of the position of the element
in the cluster.  (b) The position of an element at one
time step as a result of the force experienced at the previous time
step as a function of the
position.
(For the calculation, see the text in Section 6.)
The average strength of the force and its fluctuation were computed
over $10^4$ steps, after discarding the initial $5000$ steps, and
averaged over $1000$~samples, starting from random initial
conditions.
Here, $a=f1.8, \varepsilon=0.15,  N=100, {R}=10.0, F=0.1, L=100.0$, and 
$d=\frac{R}{2}$.}
\label{fig:potential.intermittent}
\end{figure}

To answer this question, we recall the collective attraction phenomena
in a cluster investigated in the previous section. In Section~6, it
was seen that  in the desynchronized phase, when we plot the direction
and the strength of force acting on an element as a function of the
position of the elements in the cluster, we find that there is a
collective attraction of elements that acts to form a cluster. Hence,
in this case, there exists no interaction among clusters and no
diffusion of elements.
We also calculated the force and its fluctuation in the present case.
In Figure~\ref{fig:potential.intermittent}(a), the average force experienced by an element is plotted as a function of its position,
and the position at one time step as a result of the force experienced at the previous time step is
plotted in Figure~\ref{fig:potential.intermittent}(b).  As shown,
there is an average attractive force exerted an element around the
center of the cluster, and there is a region in which the fluctuation
of the force is not large enough to change its sign.  The position at
one time step, as a function of that of at the previous time step does
not cross over the diagonal line $y=x$, except the center.  This
indicates the elements remain in a cluster for a very
long time.  We should note that in the present case, 
it is the interactions among the neighboring clusters that causes the
structure to be stable.
On the other hand, there are some neutral
zones, where the force is nearly zero (and the next position is
around the $y=x$ line) at positions larger than about 1.0.  This
leads to the intermittent switch of elements among neighboring
clusters.

\section{Summary and Discussion}

In this paper we have proposed an abstract model, a coupled map gas,
in order to study coupled dynamical systems whose couplings change in
time in a manner that depends on the states of the system elements and
their interactions.  For this purpose, we introduced ``motility'' in
space for the elements in our model, which represents a combination
and extension of CML and GCM systems. In the present model, active
elements with internal dynamics move through space and as determined
by their interactions with the other elements. The internal state of
each element is determined by these interactions as well as by its own
intrinsic dynamics.  As a result of their motion in space, the
couplings among elements can change in time.

The present model exhibits a variety of phenomena.  One of the most
characteristic types of behavior of this system is the formation of
clusters in space.  These clusters are not isolated, but rather
interact with each other.  As a result of such interactions, clusters
exchange elements.  The formation and collapse of clusters occur
repeatedly.  Depending on the parameter values, clusters can form a
lattice structure.  In such a case, the exchange of elements between
clusters is only intermittent.

The above mentioned phenomena were studied here from the viewpoint of
the interplay between the internal and external dynamics. As studied
in Section~7, the clusters form a stable lattice structure, in which
the internal oscillations of the elements in neighboring clusters are
out of phase with each other. Thus, there are two kinds of elements,
whose internal states have one of the two different phases. However,
this formation of lattice structure is not a simple pattern formation
of the ordering of two kind of elements, as is indicated by the
intermittent diffusion of elements. On the one hand, the internal
dynamics of an element depend  on its position in the structure. As a
result, two kinds of elements with different types of internal
dynamics emerge. On the other hand, the existence of two kinds of
elements leads to the formation of lattice structure. Some balance
between the effects of the dynamics of spatial degrees of freedom and
the dynamics of internal degrees of freedom is essential to the
formation of stable structure here.

As the coupling strength is increased or the nonlinear parameter of
the elements is decreased, there is an increasing tendency toward the
synchronization.  In this case, elements in a cluster can form
synchronized groups with oscillations of different phases. As a
result,  the lattice structure becomes destabilized, and the formation
and collapse of clusters take place repeatedly. Conversely, as the
coupling strength is decreased or the nonlinear parameter is
increased, the dynamics of each element become too chaotic to allow
the formation of groups with correlated phases of oscillation. In this
case too, lattice structure is destabilized, and again clusters form
and collapse repeatedly.

In this paper we have discussed a novel type of structure formation
and dynamics in a system of interacting motile elements with internal
dynamics.  Often, this structure is not rigid and is flexible, due to
the interplay between the internal dynamics and the interactions. 
Structure formation and collective dynamics of active elements are
widely seen in a biological systems as well as in a biological-type
physico-chemical systems.  The presently studied coupled map gas
system may be too simple and abstract to capture all the complexity of
such biological systems.  Still, due to the universality in the class
of systems possessing interaction motile elements of the behavior we
have investigated, we expect that the present study will sheds a new
light on the understanding of collective dynamics in biological
systems.

The authors would like to thank S. Sasa, T. Ikegami, M. Sano and 
A. Mikhailov for useful discussions.  
This research was supported by
Grants-in-Aid for Scientific
Research from the Ministry of Education, Culture, Sports,
Science and Technology of Japan (11CE2006).
T.~S. gratefully acknowledges the support from the Alexander von
Humboldt Foundation~(Germany).


\end{document}